# DeepVox and SAVE-CT: a contrast- and dose-independent 3D deep learning approach for thoracic aorta segmentation and aneurysm prediction using computed tomography scans

Matheus del-Valle[a,*], Lariza Laura de Oliveira[a], Henrique Cursino Vieira[a], Henrique Min Ho Lee[b], Lucas Lembrança Pinheiro[c], Maria Fernanda Portugal[c], Newton Shydeo Brandão Miyoshi[a], Nelson Wolosker[c]

[a] Health Innovation Techcenter, Albert Einstein Israeli Institute of Education and Research, São Paulo, Brazil

[b] Department of Radiology, Albert Einstein Israeli Institute of Education and Research, São Paulo, Brazil

[c] Department of Vascular and Endovascular Surgery, Albert Einstein Israeli Institute of Education and Research, São Paulo, Brazil

* Corresponding author. E-mail address: matheusdv@gmail.com

## Abstract

Thoracic aortic aneurysm (TAA) is a fatal disease which potentially leads to dissection or rupture through progressive enlargement of the aorta. It is usually asymptomatic and screening recommendation are limited. The gold-standard evaluation is performed by computed tomography angiography (CTA) and radiologists time-consuming assessment. Scans for other indications could help on this screening, however if acquired without contrast enhancement or with low dose protocol, it can make the clinical evaluation difficult, besides increasing the scans quantity for the radiologists. In this study, it was selected 587 unique CT scans including control and TAA patients, acquired with low and standard dose protocols, with or without contrast enhancement. A novel segmentation model, DeepVox, exhibited dice score coefficients of 0.932 and 0.897 for development and test sets, respectively, with faster training speed in comparison to models reported in the literature. The novel TAA classification model, SAVE-CT, presented accuracies of 0.930 and 0.922 for development and test sets, respectively, using only the binary segmentation mask from DeepVox as input, without hand-engineered features. These two models together are a potential approach for TAA screening, as they can handle variable number of slices as input, handling thoracic and thoracoabdominal sequences, in a fully automated contrast- and dose-independent evaluation. This may assist to decrease TAA mortality and prioritize the evaluation queue of patients for radiologists.

**Keywords:** deep learning; computed tomography; thoracic aortic aneurysm; contrast; dose; segmentation.

## 1. Introduction

Thoracic Aortic Aneurysm (TAA) is a fatal disease which potentially leads to dissection or rupture through progressive enlargement of the aorta [1–4]. Among aortic diseases, TAA are the third most frequent. They can be classified according to anatomical location in ascending aortic aneurysms (the most frequent), aortic arch aneurysms, and isolated descending thoracic aortic aneurysms [1].

Several imaging methods can be used to study the thoracic aorta. In the proximal segment, evaluation using transesophageal echocardiography is frequent [5,6]. Computed Tomography (CT) and CT Angiography (CTA), however, are more advantageous techniques, as they allow a complete assessment of the aorta and reduce the evaluator influence in the vessel diameter assessment. Most methods, however, rely on the use of endovascular contrast for better differentiation between the aortic lumen and its surroundings [5,6]. In non-contrasted CT images, therefore, the aorta segmentation can be a challenging task.

Screening tools have demonstrated positive impact by decreasing Abdominal Aortic Aneurysm (AAA) mortality [7]. Yet, asymptomatic thoracic aortic imaging screening is only recommended for first-degree relatives of individuals with thoracic aortic disease or for patients with genetic mutations that predispose the aneurysm development [8]. This may fail to identify several patients, once TAA is typically asymptomatic until acute dissection or aortic rupture occur, and only around 20% of patients with TAA or dissection have a first-degree relative with a similar disease [9]. Aneurysms are often discovered incidentally on imaging studies for other indications [10]. Low dose chest CT could be used for TAA screening, as they are performed in annual lung cancer screening for patients from 50 to 80 years old who have smoking history in the past 15 years [11], which are also risk factors for TAA disease [8]. Nevertheless, besides the non-contrast enhancement, the lower radiation implies lower image quality, making the evaluation of the aorta even more difficult.

In addition to image quality and enhancement challenges, the radiology reporting is a tedious and time-consuming task, with inter-observer variabilities [12], especially geometric measures used in clinical practice, which is an oversimplified and error-prone approach [13]. Therefore, several studies developed automated algorithms and machine learning models for segmentation tasks [14–20], and aortic assessment [19–22]. Still, there is no study using a unified approach for all kind of CT: with or without contrast enhancement, low or standard dose protocol; and not dependent on hand-engineering features, such as diameter measurement, rather than the model learning by itself the features that make the scan classification possible as in a deep learning approach [23]. In this way, this study aims to develop a contrast- and dose-independent model using an end-to-end deep learning method for thoracic aorta segmentation and TAA classification.

## 2. Material and Methods

### 2.1. Dataset gathering

It was conducted a retrospective study of thoracoabdominal CT scans acquired from 2016 to 2021 at the Hospital Israelita Albert Einstein (HIAE, São Paulo, Brazil). The study was approved by the institutional review board of Hospital Israelita Albert Einstein (Project 4531-20) and by the National Commission for Ethics in Research of the National Health Council of the Ministry of Health (CAAE 44951021.8.0000.0071).

Scans were acquired using several CT equipment of the hospital from three manufacturers: GE HealthCare (Illinois, USA), Siemens Healthineers (Munich, Germany) and Canon Medical Systems (Tochigi, Japan). It was selected unique patients who underwent CT scan using standard or low radiation dose protocol, with or without contrast enhancement, presenting or not a TAA. Exclusion criteria included the presence of CT artefact due movement or medical metal devices, and any known aneurysm treatment, such as endografts, valve grafts and active aortic dissections.

After data selection, 587 unique scans were gathered, anonymized and split in five groups: LD (Low Dose), SD (Standard Dose), CTA (CT Angiography), AN (Aneurysm), and ANNC (Aneurysm No-Contrast). **Table 1** presents the dataset groups distribution. Scans characteristics were extracted using the DICOM tags.

**Table 1**

Dataset distribution by group.

| Group | Scans | Dose Protocol | Contrast Enhancement | TAA Presence |
|---|---|---|---|---|
| LD | 150 | Low | No | No |
| SD | 150 | Standard | No | No |
| CTA | 150 | Standard | Yes | No |
| AN | 119 | Standard | Yes | Yes |
| ANNC | 18 | Standard | No | Yes |

*2.2. Data annotation*

Manual semantic segmentation was performed by three expert physicians (one radiologist and two vascular surgeons) using the 3D Slicer software [24]. Thoracic aortas were annotated slice-by-slice through the axial plane, while also considering the sagittal and coronal planes for structures identification and corrections, leading to binary masks as outputs. The aorta initial point was considered as just above the aortic valve or as an arbitrary point 2.5 cm proximal to the brachiocephalic trunk when the valve identification was not possible. The final point was just above diaphragmatic hiatus or the diaphragmatic cupula if the hiatus was not possible to be identified. Arterial wall regions were included in the annotation.

Sixteen random LD scans were annotated in triplicate to perform an inter-observer evaluation using the Dice Score Coefficient (DSC) metric. A voting system was applied to these scans to generate a single segmentation mask, where each pixel must had at least two annotators overlay, otherwise it was discarded. The remaining scans were distributed without overlay and keeping the group representativity uniform for all the annotators.

Segmentations were revised by a researcher with medical segmentation experience and saved as NRRD (Nearly Raw Raster Data) file format. These final masks were considered as the Ground Truth (GT) segmentation for the deep learning models. GT binary classes (control or TAA) were extracted from the medical reports and confirmed by medical experts' scans assessment for the classification model.

*2.3. Data preprocessing*

The Preprocessing was defined according to the deep learning models, hardware and scans qualitative analysis. The image shape was set as the maximum shape without losing information from regions near the aorta or overloading the 32 GB graphics card used for the segmentation models. The following preprocessing steps were applied for image segmentation:

- Hounsfield Unit (HU) conversion: the raw intensities of the scans were converted to HU using the slope and interception DICOM tag values.
- Positioning: patient positioning variations were standardized by setting "head first-supine" (HFS) orientation for all the scans and masks.
- Windowing: a soft tissue window was applied to the scans using a level of 50 and width of 400 HU [25].
- Resampling: scans and masks were transformed to an isomorphic resolution of 2x2x3 mm (X pixel size x Y pixel size x thickness). This was performed using the nearest-neighbor interpolation to preserve the image-binary mask spatial relation.
- Crop: to keep a fixed shape for the models and remove regions too far from the aorta, scans and masks were cropped to 128-pixel size (X and Y). The Z length was not changed in this step.
- Z-Reshape: for fixed Z length model, scans and masks were transformed to 128 slices using nearest-neighbor interpolation.
- NRRD: all files were saved as NRRD file format to optimize the compression and loading.

In this way, inputs for segmentation were isomorphic images of 128x128xVariable for variable Z models; and 128x128x128 for fixed Z models. All images had a single channel: the windowed HU intensity. The TAA prediction was achieved using the output masks prediction. A Z-trimming step was applied to the masks to remove the excess background. Final predicted masks dataset was analyzed for possible disconnected small segments due to wrong deep learning segmentation, which were removed using the pixel connectivity labeling [26] and a fixed removal threshold of 5% of the total mask volume.

*2.4. Deep learning*

Three deep learning models were trained and evaluated to execute the thoracic aorta segmentation, while one model was developed to perform the TAA prediction.

2.4.1. Thoracic aorta segmentation

The automatic semantic segmentation was accomplished by one model with fixed Z length (fixed slices number): 3D U-Net [27]; and two models with variable Z: DeepAAA [19] and DeepVox. The 3D U-Net is a volumetric version of U-Net [28], one of the most established and used models for segmentation in the literature. DeepAAA is a variant of 3D U-Net, where the pooling kernels have size of 3x3x1 to preserve the Z-dimension, hence receiving scans and masks with variable slices number.

The novel model DeepVox is a conditional Generative Adversarial Network (cGAN) [29–31] in a 3D approach with variable Z. The generator (**Fig. 1**) and the discriminator (**Fig. 2**) were developed based on DeepAAA and Vox2Vox [32] – 3D cGAN to segmentate brain tumors, where the generator has a U-Net and Res-Net [33,34] style, and the discriminator is a PatchGAN [35]. The losses were set similar to Vox2Vox: discriminator loss as the sum of the L2 error; generator loss as discriminator loss plus five times the hybrid focal loss [36], instead of dice loss as in Vox2Vox. All convolutional layers were created using He Normal kernel initialization [37] and zero padding.

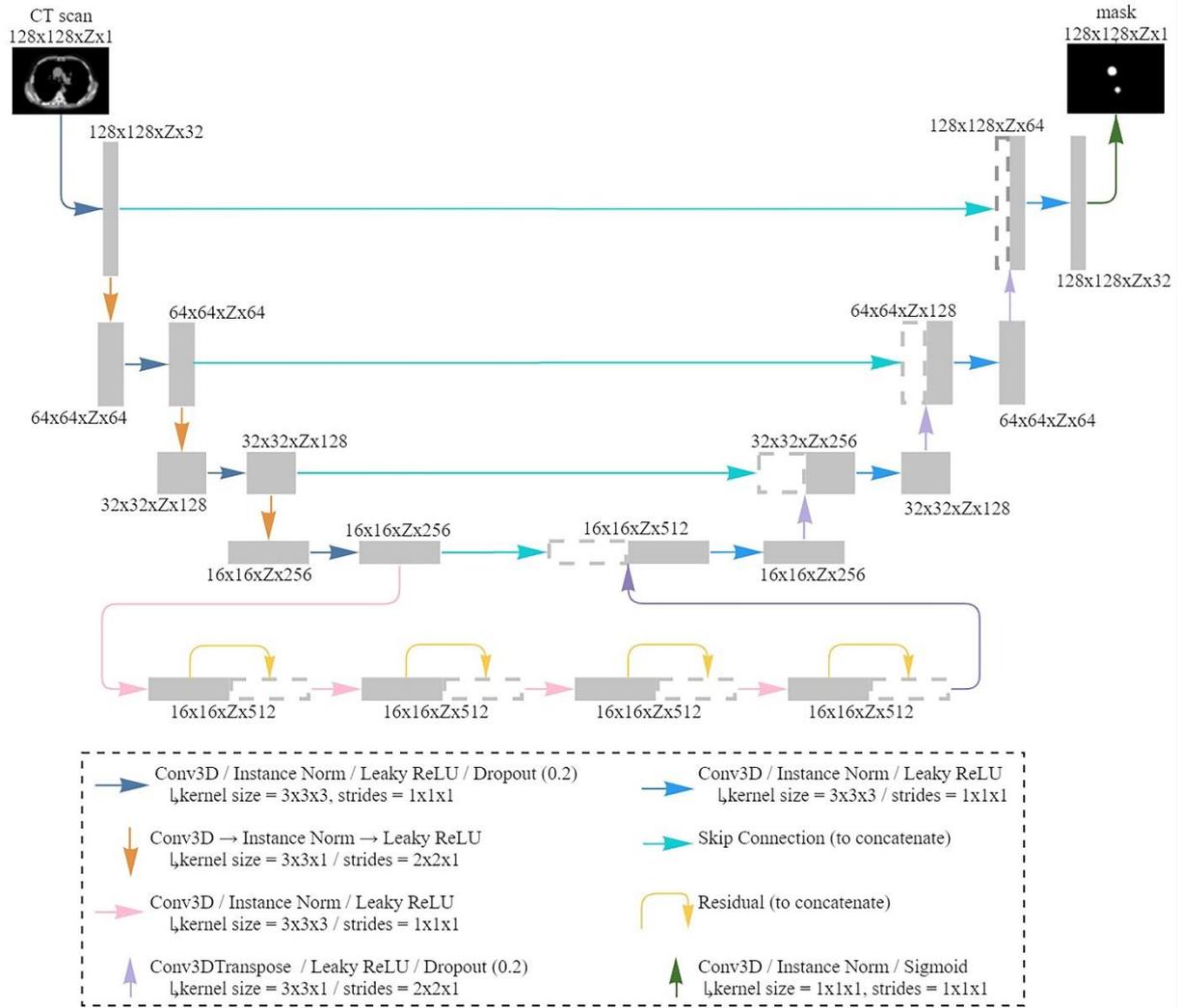

**Fig. 1.** Generator architecture of DeepVox. The "Z" in the shape means variable dimension.

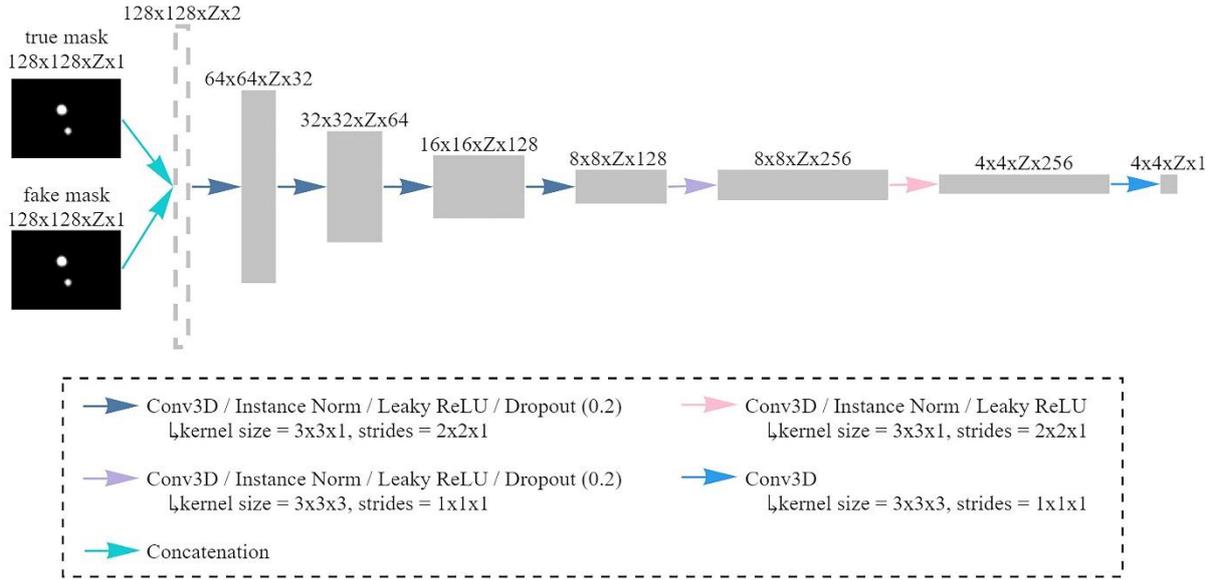

**Fig. 2.** Discriminator architecture of DeepVox. The "Z" in the shape means variable dimension.

The 16 LD scans annotated in triplicated were used to evaluate the voting impact. Three DeepAAA models were trained in a 4-fold cross-validation with the following scenarios of train/dev (development) sets split: (i) 12/4 with voting; (ii) 12/4 without voting; (ii) 4/12 with voting. Masks without voting were randomly distributed from each annotator. Statistical comparison of the scenarios was performed by Friedman test with Nemenyi post-hoc test [38] with a significance level of 5%.

An initial sampling of 117 scans, representative of the final dataset distribution, was used to choose the best approach of the data organization for the segmentation process. Three DeepAAA models were trained by a stratified 4-fold cross-validation, where their difference was the input dataset: (i) contrast enhanced scans; (ii) non-enhanced scans; (iii) all scans. 3D U-Net and DeepVox were also trained using approach (iii). The performance of approach (iii) and the concatenation (i) and (ii), and performances of the three models using approach (iii) were statistically compared using Friedman + Nemenyi test.

For the final training, with the complete dataset, a total of 60 scans were held-out to the test set: 15 LD, 15 SD, 15 CTA, 12 AN, and 3 ANNC; while the 527 remaining were trained by a stratified 4-fold cross-validation, resulting in 395 scans for the train set and 132 for the dev set of each fold. DeepAAA and DeepVox were trained and evaluated with the complete dataset.

All models were trained using batch size of 1 and Adam [39] optimizer with learning rate of 1e-3, Beta 1 of 0.5 and Beta 2 of 0.999. Final training was accomplished by 300 epochs and using cosine decay schedule with restarts [40], with initial learning rate of 1e-3, first decay step with the length of the training set, epochs multiplier in the decay cycle (t_mul) of 1.5, initial learning rate multiplier (m_mul) of 1.0, and minimum learning rate (alpha) of 1e-6.; while the other evaluations were performed along 50 epochs and using the reduce learning rate on plateau callback [41] with patient of 5, reduce factor of 0.5 and minimum learning rate of 1e-5. DSC, precision and sensitivity metrics were calculated to evaluate models' performance.

A data generator randomly shuffled and augmented the training data by every epoch. Augmentation involved different random transformations for the scan and mask: 3D rotation and flip; brightness change, only in scans, using power-law gamma transformation [42] with random gamma and gain between 0.9 and 1.1; and elastic deformation [27,28] with sigma of 2. Augmentation did not duplicate or increase the dataset size. All segmentation were trained and tested using a 32 GB graphics card (NVIDIA Tesla V100S).

2.4.2. Thoracic aortic aneurysm classification

A deep learning model was developed to predict the TAA directly and solely from the segmentation (binary mask). The novel model, named here by SAVE-CT (Screening of Aneurysm on Variable Exams of Computed Tomography), is based on DeepVox discriminator, added one Conv3D to each convolutional layer. The output layer was changed to a 3D global average pooling and a single last neuron with sigmoid activation for binary classification. **Fig. 3** illustrates the model architecture.

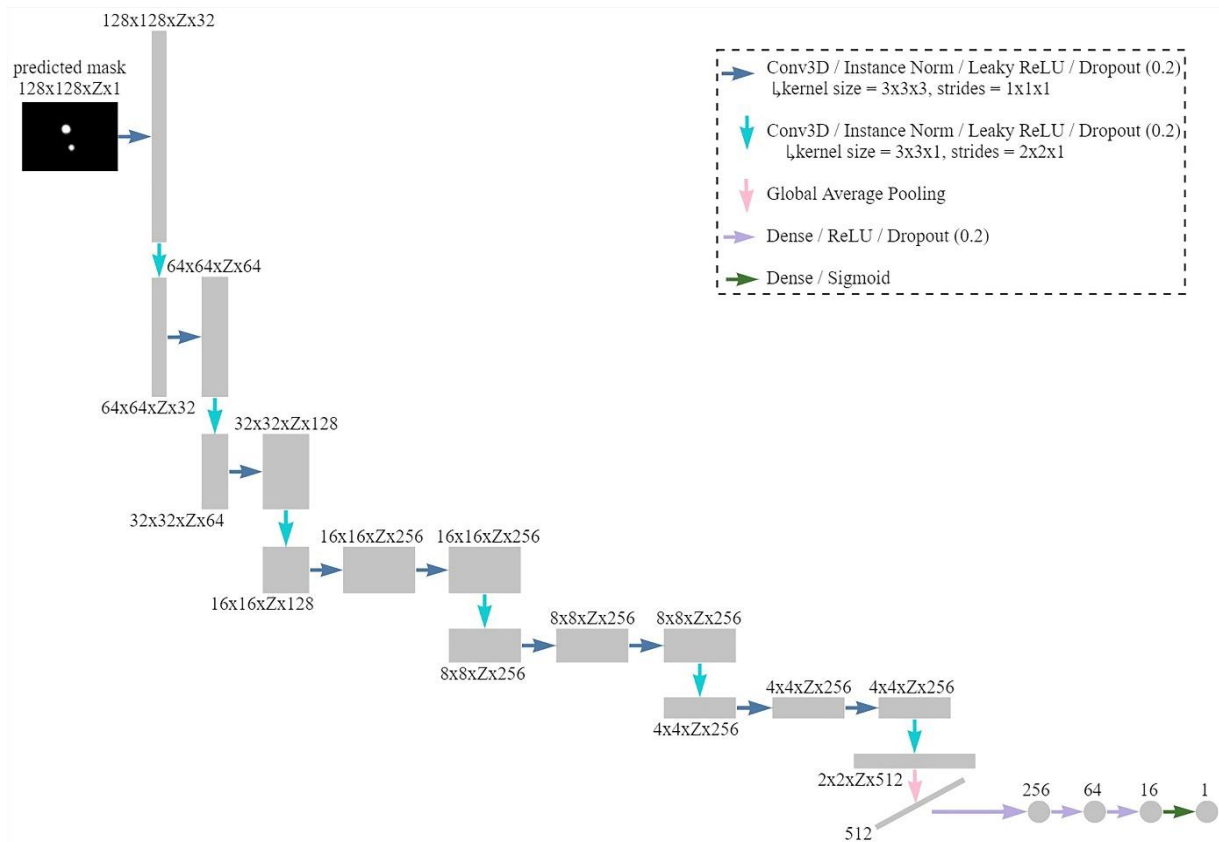

**Fig. 3.** SAVE-CT architecture. The "Z" in the shape means variable dimension.

Masks generated from each dev set of DeepVox were used for the training in a stratified 10-fold cross-validation, where a class balance was executed by downsampling the control scans (class zero) to meet the same quantity of AN and ANNC (class one). The 60 test scans generated masks were also separated to the test set of the classification.

SAVE-CT model was trained using binary cross-entropy loss with Adam optimization – learning rate of 1e-4, Beta 1 of 0.5 and Beta 2 of 0.999. It was trained for 50 epochs, with batch size of one, and RLROP – patient of 5, reduce factor of 0.5 and minimum learning rate of 1e-5. Performance was assessed by binary accuracy, precision, sensitivity and specificity metrics, and by a 3D adaptation of the Gradient-weighted Class Activation Mapping (Grad-CAM) [43] analysis. A 3D min-max normalization was applied to the Grad-CAM.

SAVE-CT training and testing phases were accomplished using an 8 GB graphics card (GeForce GTX 1080). This whole study was performed by in house algorithms in Python, mainly using Tensorflow and Keras libraries.

## 3. Results and Discussion

The **Table 2** presents the dataset characteristics. Scans resolution (pixel size and thickness) were similar between groups due to standardized CT protocols at the HIAE. LD and SD demonstrated lower slices quantity and standard deviation, once these groups contained more thoracic scans than the others, composed mainly by thoracoabdominal scans, especially LD which is mainly a lung cancer screening protocol [44]. AN and ANNC groups presented the greater difference between men and women distribution, and greater age and weight mean than the other groups. These characteristics may be related to reported findings regarding TAA: predominant incidence in men [45]; onset aged around 60 years [46]; and positive correlation with weight gain [47].

**Table 2**

Dataset characteristics by group. Mean ± standard deviation.

| Group | Pixel Size (mm) | Thickness (mm) | Slices Quantity | Male / Female | Age (years) | Height (m) | Weight (kg) |
|---|---|---|---|---|---|---|---|
| LD | 0.69 ± 0.07 | 1.00 ± 0.00 | 332 ± 29 | 72 / 78 | 61 ± 4 | 1.66 ± 0.09 | 73.6 ± 15.9 |
| SD | 0.77 ± 0.10 | 0.98 ± 0.09 | 352 ± 50 | 96 / 54 | 43 ± 18 | 1.71 ± 0.09 | 72.7 ± 16.9 |
| CTA | 0.77 ± 0.09 | 1.06 ± 0.19 | 408 ± 122 | 89 / 61 | 56 ± 15 | 1.70 ± 0.10 | 74.8 ± 15.4 |
| AN | 0.76 ± 0.09 | 0.99 ± 0.19 | 426 ± 135 | 80 / 39 | 66 ± 13 | 1.70 ± 0.09 | 77.2 ± 17.7 |
| ANNC | 0.76 ± 0.10 | 0.92 ± 0.15 | 409 ± 81 | 13 / 5 | 70 ± 18 | 1.73 ± 0.09 | 83.7 ± 17.9 |

*3.1. Thoracic Aorta Segmentation*

Inter-observer and model's dev set DSC for the 16 LD scans annotated in triplicate are described in **Table 3**. Inter-observer metrics and approaches (i) and (ii) presented close DSC values, indicating that the variability of the models is similar to that of the experts. Several tasks regarding aorta annotation are reported to present considerably inter-observer variability [48–53]. LD scans were chosen for this assessment as they present the hardest manual segmentation due to the lack of structures enhancement by intravenous contrast and lower image quality caused by the low dose of radiation [54].

**Table 3**

Dice Score Coefficient (DSC) for inter-observer masks comparison and for the dev sets of segmentation models trained in scenarios (i), (ii) and (iii). Mean ± standard deviation.

| Scenario | DSC |
| --- | --- |
| Inter-observer | 0.857 ± 0.037 |
| (i) 12/4 with voting | 0.876 ± 0.020 |
| (ii) 12/4 without voting | 0.855 ± 0.034 |
| (iii) 4/12 with voting | 0.784 ± 0.042 |

Scenario (i) showed the best DSC, indicating better mask quality. Still, there was no statistical between scenarios (i) and (ii). Several experts' segmentation masks of a same patient help to deal with inter-observer variability; however, the ground truth annotation is time-consuming and it is common to provide only one mask for each patient [55]. In addition, both scenarios presented significant statistical difference from scenario (iii), hence it is better to have more samples without voting, than less samples with voting. This indicates that the mask quality difference was not comparable to the gain of having more samples, which corroborates the fact that CNN models strongly benefit from larger medical images datasets [56].

The **Table 4** shows the DSC for dataset approaches (i), (ii), and (iii), where the values are close to each other. Using two models, from approach (i) and (ii), is an alternative for one model for all kind of data (iii). However, there was no significant statistical difference between approach (iii) and the concatenation of (i) and (ii) DSC. The mean between (i) and (ii) of 0.842 ± 0.048 demonstrates an even closer value to (iii) of 0.841 ± 0.039. These findings indicate that one single model can generalize the learning with both enhanced and non-enhanced scans.

**Table 4**

Dev sets dice score coefficient (DSC) of segmentation models trained with dataset approaches: (i) contrast enhanced scans; (ii) non-enhanced scans; (iii) all scans. Mean ± standard deviation.

| Approach | DSC |
| --- | --- |
| (i) contrast enhanced | 0.836 ± 0.039 |
| (ii) non-enhanced | 0.849 ± 0.060 |
| (iii) all scans | 0.841 ± 0.039 |

Performance of segmentation models trained using the initial sampling is shown in **Table 5**. Metrics were close between the models, where 3D U-Net presented the lowest DSC and precision, while DeepVox showed the best DSC and sensitivity. Although comparable with the others, 3D U-Net is a fixed Z model, which demands flattened scans, specially thoracoabdominal ones. This results in masks with lower quality, thus this model should be considered for approaches that the aorta limits are well defined, which is not the case of this study, where scans from variable lengths are included. Variable Z axis enables DeepAAA and DeepVox to receive both thoracic and thoracoabdominal exams, identifying where the thoracic aorta ends and not proceeding the segmentation to the abdominal region.

**Table 5**

Dev set performance comparison of segmentation models trained using initial sampling – approach (iii) all scans. Mean ± standard deviation.

| Model | DSC | Precision | Sensitivity | Time per Epoch (min) |
|---|---|---|---|---|
| 3D U-Net | 0.826 ± 0.041 | 0.869 ± 0.047 | 0.790 ± 0.103 | 8.47 |
| DeepAAA | 0.841 ± 0.040 | 0.902 ± 0.034 | 0.775 ± 0.099 | 7.95 |
| DeepVox | 0.853 ± 0.032 | 0.896 ± 0.046 | 0.797 ± 0.082 | 2.40 |

DeepVox conception was based on the Vox2Vox cGAN with variable Z from DeepAAA. Yet, the performance was not satisfactory, leading to several other changes, as adding a VGG-11 based model to the discriminator. Afterwards, the model was able to achieve the best DSC and a training time per epoch more than 3x lower than the others. Lighter models are preferred for future studies, as they enhance the assessments and enable a larger number of hyperparameters being tested in a shorter time. This difference may be even more pronounced if considered a real prediction task, where only the generator model is used, since the discriminator is useful only for training tasks, decreasing the number of parameters to be calculated. Therefore, DeepVox demonstrates a better applicability than 3D U-Net and DeepAAA.

The **Table 6** presents the performance of DeepAAA and DeepVox with the complete dataset. 3D U-Net was not considered because its previous results were unsatisfactory. DeepVox presented better results for all the metrics.

**Table 6**

Dev and test sets performance comparison of segmentation models trained using final dataset. Mean ± standard deviation.

| Model | Set | DSC | Precision | Sensitivity |
|---|---|---|---|---|
| DeepAAA | Dev | 0.917 ± 0.037 | 0.958 ± 0.031 | 0.898 ± 0.057 |
| DeepAAA | Test | 0.876 ± 0.039 | 0.901 ± 0.036 | 0.865 ± 0.058 |
| DeepVox | Dev | 0.932 ± 0.028 | 0.966 ± 0.033 | 0.911 ± 0.040 |
| DeepVox | Test | 0.897 ± 0.035 | 0.923 ± 0.029 | 0.887 ± 0.049 |

DSC achieved by DeepVox is comparable to the values presented in the literature. Lu et al, 2019 [19] exhibited a DSC of 0.90 by training the DeepAAA model on 321 exams (223 unique patients), where 48% were contrast-enhanced and 77% containing abdominal aortic aneurysm (AAA), and testing on 57 unique exams, where 51% were contrast-enhanced and 51% contained AAA. Comelli et al., 2020 [57] performed a 5-fold cross-validation with 72 patients, using contrast-enhanced ECG-gated CT exams. UNet and ENet [58] models' segmentations of the ascending TAA showed a DSC of 0.91. Adopting an anatomy label maps algorithm, Xie et al., 2014 [59] reported a DSC of 0.93 using low-dose non-contrast exams from VIA-ELCAP database and non-contrast exams from LIDC database. Macruz et al., 2022 [20] showed a DSC of 0.92 using 3D U-Net in scans with and without contrast enhancement. Furthermore, DeepVox is the most generalized model so far, the only one dealing with low and standard dose, contrast and non-contrast enhanced, healthy and aneurysmal, thoracic and thoracoabdominal scans, and considering the whole thoracic aorta (ascending, arch and descending regions).

Although the DSC is an established metric to assess the model performance, it often exhibits high precision, but low sensitivity when its loss version (dice loss, defined by 1-DSC) is applied to class imbalanced problems, such as the thoracic aorta, which is a small volume compared to the background (whole scan). Several approaches which aim to improve the imbalanced performance were evaluated, where the optimal function selected was the Hybrid Focal Loss, a combination of the Focal Tversky Loss and the Focal Loss. Thereby, it was possible to balance the precision and sensitivity performance, as the 0.923 ± 0.029 and 0.887 ± 0.049, respectively, presented by the DeepVox test set.

Models with a 2D approach, for slice-by-slice segmentation, were also evaluated, such as Pix2Pix [60], U-Net and its variants [58,61–63]. Yet, they did not perform satisfactorily, exhibiting metrics considerably lower than 3D models and a high impact of flawed masks, with several "holes" in them. Even using a postprocessing filling hole algorithm [64] was not able to improve the segmentation quality enough.

*3.2. Thoracic Aortic Aneurysm Classification*

The **Table 7** presents SAVE-CT TAA prediction metrics.

**Table 7**

Dev and test sets performance of SAVE-CT TAA classification. Mean ± standard deviation.

| Set | Accuracy | Precision | Sensitivity | Specificity | F1 Score |
|---|---|---|---|---|---|
| Dev | 0.930 ± 0.042 | 0.928 ± 0.042 | 0.933 ± 0.057 | 0.926 ± 0.048 | 0.928 ± 0.047 |
| Test | 0.922 ± 0.038 | 0.782 ± 0.043 | 0.955 ± 0.060 | 0.911 ± 0.035 | 0.857 ± 0.043 |

SAVE-CT performance is better when compared to studies in the literature. Macruz et al., 2022 [20] reported an accuracy of 70 to 86%, sensitivity of 88 to 93%, and specificity of 57 to 81% when classifying TAA. Several other studies also present machine learning evaluations of AAA [65], where Shum et al, 2011 [66] showed an average accuracy of 86.6% for AAA classification using a decision tree-based model. Beyond the better accuracy, the considerably higher specificity of SAVE-CT in comparison to Macruz et al., 2022 demonstrates it as better approach to deal with the real imbalance distribution of TAA, where the estimated prevalence is between 0.16 and 0.34% [67]. This means that a low specificity may result in a high number of false positives, not being useful as a screening tool.

Besides the better performance metrics, SAVE-CT is the only model so far using a full deep learning approach and without relying on absolute diameter measurements or other radiomics approaches, extracting morphological features automatically. This could lead to other studies evaluating the bias of diameter measure tools as an aneurysm screening in comparison to automated deep learning methods as SAVE-CT.

**Fig. *4*** shows a representative image of the Grad-CAM evaluation. The normalization of the plot to allow a single scale for all the slices made it possible to understand each slice influence in the classification. The TAA presented in the arch region contributed the most to the classification, denoted by the red mapping of the mask. In contrast, ascending and descending slices had lower impact, as no aneurysm was present, and the background had minimal influence as it was of only zeroes. This finding indicates SAVE-CT correct identification of the aneurysmal region.

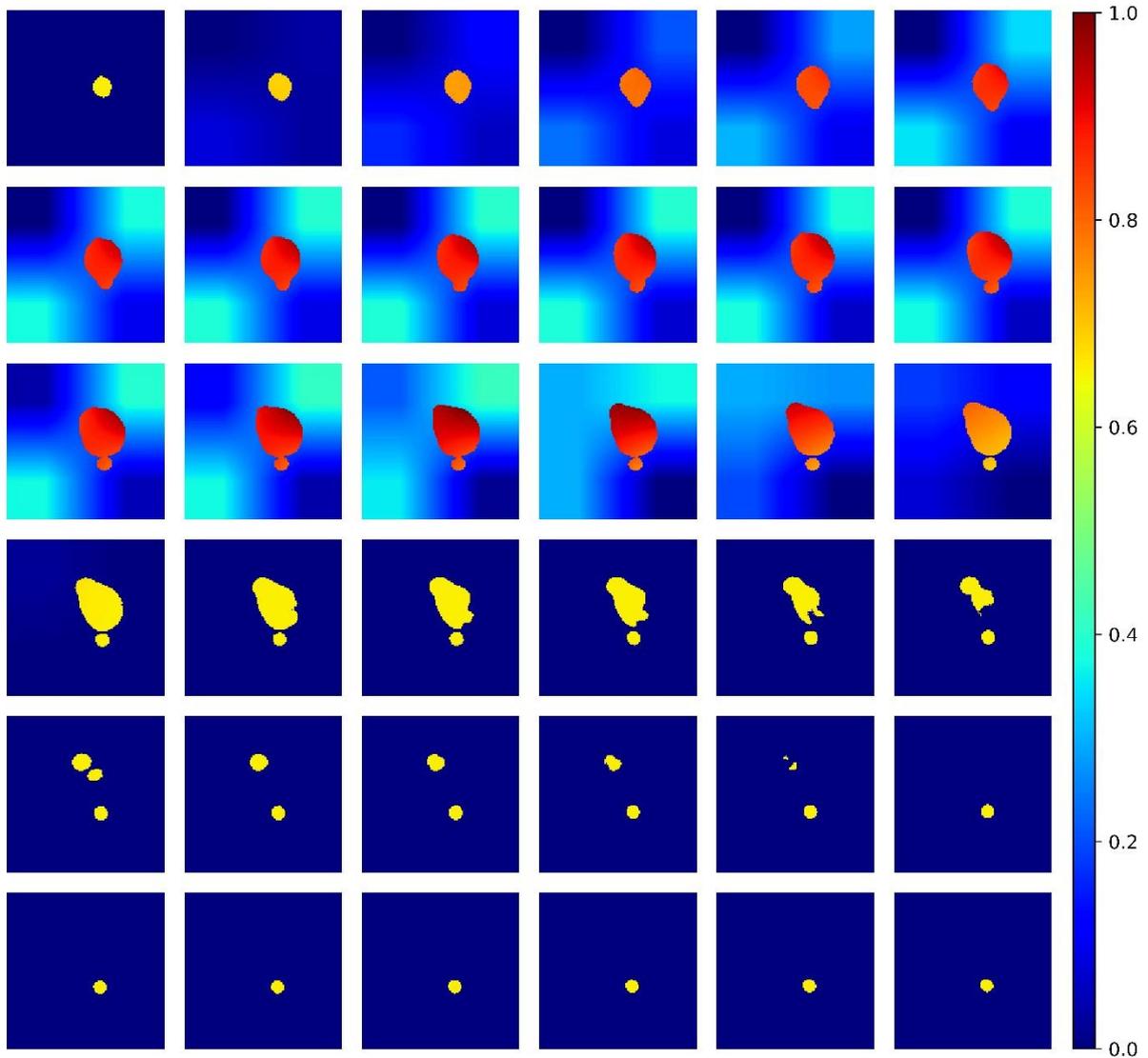

**Fig. 4.** Representative image of the Gradient-weighted Class Activation Mapping (Grad-CAM) from a AN (aneurysm) scan. The plot considered 32 slices near the aneurysm region,

The variable depth feature of the classification model adds the ability to handle any number of slices, without overly compressing the Z axis, in which an aneurysm region could be eliminated. Moreover, the ability of the segmentation model to deal with both thoracoabdominal and thoracic scans has been maintained in the classification.

The average training time of SAVE-CT was 1.4 min per epoch. Since it depends on binary masks, already resampled and trimmed, it was possible to execute a fast training with a simpler hardware in comparison to the segmentation models, with a quarter of total GPU memory.

The limitations of this study are mainly related to its retrospective approach. Although scans were from more than one CT manufacture, it represents the standardized quality scans from Hospital Albert Einstein and from Brazil, demanding validations and fine tuning with patients from several hospitals and countries. In addition, a clinical trial must be carried out to assess the models in a real environment, with representative distribution of TAA and statistical analysis of impact of using or not this AI screening tool.

## 4. Conclusion

The novel DeepVox model demonstrated comparable metrics to the literature for thoracic aorta segmentation, with a dice score and training speed improvement, while accepting any kind of computed tomography scan: with or without contrast enhancement, acquired with low or standard dose protocol. SAVE-CT showed the best metrics for thoracic aortic aneurysm classification in comparison to other studies, being the first model to use only the binary segmentation mask as input, thus not relying on hand-engineered features.

DeepVox and SAVE-CT together allows scans with variable number of slices as input, which is important to handle thoracic and thoracoabdominal sequences, providing a fully automated evaluation for several CT scan protocols. This indicates a potential approach for thoracic aortic aneurysm screening, which may assist to decrease its mortality and prioritize the evaluation queue of patients for radiologists.

**Acknowledgements**

The authors would like to thank GE HealthCare Brazil and USA for their support during the project.

**Conflict of interest**

The authors do not have any conflicts of interest to disclose.

**Funding**

This work was funded by GE HealthCare Brazil, through the "Brazilian Informatic Law" – Law No. 8,248 of October 23rd, 1991.

# References

[1] Rogers IS, Massaro JM, Truong QA, Mahabadi AA, Kriegel MF, Fox CS, et al. Distribution, Determinants, and Normal Reference Values of Thoracic and Abdominal Aortic Diameters by Computed Tomography (from the Framingham Heart Study). Am J Cardiol 2013;111:1510–6. https://doi.org/10.1016/j.amjcard.2013.01.306.

[2] Isselbacher EM, Lino Cardenas CL, Lindsay ME. Hereditary Influence in Thoracic Aortic Aneurysm and Dissection. Circulation 2016;133:2516–28. https://doi.org/10.1161/CIRCULATIONAHA.116.009762.

[3] Laukka D, Pan E, Fordell T, Alpay K, Rahi M, Hirvonen J, et al. Prevalence of thoracic aortic aneurysms and dilatations in patients with intracranial aneurysms. J Vasc Surg 2019;70:1801–8. https://doi.org/10.1016/j.jvs.2019.01.066.

[4] Faggion Vinholo T, Brownstein AJ, Ziganshin BA, Zafar MA, Kuivaniemi H, Body SC, et al. Genes Associated with Thoracic Aortic Aneurysm and Dissection: 2019 Update and Clinical Implications. AORTA 2019;07:099–107. https://doi.org/10.1055/s-0039-3400233.

[5] Tamborini G, Galli CA, Maltagliati A, Andreini D, Pontone G, Quaglia C, et al. Comparison of Feasibility and Accuracy of Transthoracic Echocardiography Versus Computed Tomography in Patients With Known Ascending Aortic Aneurysm. Am J Cardiol 2006;98:966–9. https://doi.org/10.1016/j.amjcard.2006.04.043.

[6] Evangelista A, Flachskampf FA, Erbel R, Antonini-Canterin F, Vlachopoulos C, Rocchi G, et al. Echocardiography in aortic diseases: EAE recommendations for clinical practice. Eur J Echocardiogr 2010;11:645–58. https://doi.org/10.1093/ejechocard/jeq056.

[7] Guirguis-Blake JM, Beil TL, Senger CA, Whitlock EP. Ultrasonography Screening for Abdominal Aortic Aneurysms: A Systematic Evidence Review for the U.S. Preventive Services Task Force. Ann Intern Med 2014;160:321. https://doi.org/10.7326/M13-1844.

[8] Isselbacher EM, Preventza O, Hamilton Black J, Augoustides JG, Beck AW, Bolen MA, et al. 2022 ACC/AHA Guideline for the Diagnosis and Management of Aortic Disease: A Report of the American Heart Association/American College of Cardiology Joint Committee on Clinical Practice Guidelines. Circulation 2022;146. https://doi.org/10.1161/CIR.0000000000001106.

[9] Hannuksela M, Stattin E-L, Johansson B, Carlberg B. Screening for Familial Thoracic

Aortic Aneurysms with Aortic Imaging Does Not Detect All Potential Carriers of the Disease. AORTA 2015;03:1–8. https://doi.org/10.12945/j.aorta.2015.14-052.

[10] Holloway BJ, Rosewarne D, Jones RG. Imaging of thoracic aortic disease. Br J Radiol 2011;84:S338–54. https://doi.org/10.1259/bjr/30655825.

[11] Krist AH, Davidson KW, Mangione CM, Barry MJ, Cabana M, Caughey AB, et al. Screening for Lung Cancer. JAMA 2021;325:962. https://doi.org/10.1001/jama.2021.1117.

[12] Rueckel J, Reidler P, Fink N, Sperl J, Geyer T, Fabritius MP, et al. Artificial intelligence assistance improves reporting efficiency of thoracic aortic aneurysm CT follow-up. Eur J Radiol 2021;134:109424. https://doi.org/10.1016/j.ejrad.2020.109424.

[13] Wieben O. Improved CT Surveillance of Thoracic Aortic Aneurysm Growth. Radiology 2022;302:226–7. https://doi.org/10.1148/radiol.2021212122.

[14] Zhong J, Bian Z, Hatt CR, Burris NS. Segmentation of the thoracic aorta using an attention-gated u-net. In: Drukker K, Mazurowski MA, editors. Med. Imaging 2021 Comput. Diagnosis, SPIE; 2021, p. 19. https://doi.org/10.1117/12.2581947.

[15] Kurugol S, San Jose Estepar R, Ross J, Washko GR. Aorta segmentation with a 3D level set approach and quantification of aortic calcifications in non-contrast chest CT. 2012 Annu. Int. Conf. IEEE Eng. Med. Biol. Soc., IEEE; 2012, p. 2343–6. https://doi.org/10.1109/EMBC.2012.6346433.

[16] Noothout JMH, de Vos BD, Wolterink JM, Isgum I. Automatic Segmentation of Thoracic Aorta Segments in Low-Dose Chest CT 2018. https://doi.org/10.1117/12.2293114.

[17] Moccia S, De Momi E, El Hadji S, Mattos LS. Blood vessel segmentation algorithms — Review of methods, datasets and evaluation metrics. Comput Methods Programs Biomed 2018;158:71–91. https://doi.org/10.1016/j.cmpb.2018.02.001.

[18] SN K, Fred A L, S M, H AK, Varghese P S. A voyage on medical image segmentation algorithms. Biomed Res 2018. https://doi.org/10.4066/biomedicalresearch.29-16-1785.

[19] Lu JT, Brooks R, Hahn S, Chen J, Buch V, Kotecha G, et al. DeepAAA: Clinically Applicable and Generalizable Detection of Abdominal Aortic Aneurysm Using Deep Learning. Lect. Notes Comput. Sci. (including Subser. Lect. Notes Artif. Intell. Lect. Notes Bioinformatics), vol. 11765 LNCS, Springer; 2019, p. 723–31. https://doi.org/10.1007/978-3-030-32245-8_80.

[20] Macruz FB de C, Lu C, Strout J, Takigami A, Brooks R, Doyle S, et al. Quantification of the Thoracic Aorta and Detection of Aneurysm at CT: Development and Validation

of a Fully Automatic Methodology. Radiol Artif Intell 2022;4. https://doi.org/10.1148/ryai.210076.

[21] Kontopodis N, Klontzas M, Tzirakis K, Charalambous S, Marias K, Tsetis D, et al. Prediction of abdominal aortic aneurysm growth by artificial intelligence taking into account clinical, biologic, morphologic, and biomechanical variables. Vascular 2022:170853812210778. https://doi.org/10.1177/17085381221077821.

[22] Lindquist Liljeqvist M, Bogdanovic M, Siika A, Gasser TC, Hultgren R, Roy J. Geometric and biomechanical modeling aided by machine learning improves the prediction of growth and rupture of small abdominal aortic aneurysms. Sci Rep 2021;11:18040. https://doi.org/10.1038/s41598-021-96512-3.

[23] Cheng PM, Montagnon E, Yamashita R, Pan I, Cadrin-Chênevert A, Perdigón Romero F, et al. Deep Learning: An Update for Radiologists. RadioGraphics 2021;41:1427–45. https://doi.org/10.1148/rg.2021200210.

[24] Fedorov A, Beichel R, Kalpathy-Cramer J, Finet J, Fillion-Robin J-C, Pujol S, et al. 3D Slicer as an image computing platform for the Quantitative Imaging Network. Magn Reson Imaging 2012;30:1323–41. https://doi.org/10.1016/j.mri.2012.05.001.

[25] Whalen B. The Slice Is Right (An Exercise in CT Windowing). Can J Med Radiat Technol 2003;34:5–10. https://doi.org/10.1016/S0820-5930(09)60033-5.

[26] Wu K, Otoo E, Shoshani A. Optimizing connected component labeling algorithms. In: Fitzpatrick JM, Reinhardt JM, editors., 2005, p. 1965. https://doi.org/10.1117/12.596105.

[27] Çiçek Ö, Abdulkadir A, Lienkamp SS, Brox T, Ronneberger O. 3D U-Net: Learning Dense Volumetric Segmentation from Sparse Annotation 2016.

[28] Ronneberger O, Fischer P, Brox T. U-Net: Convolutional Networks for Biomedical Image Segmentation 2015.

[29] Creswell A, White T, Dumoulin V, Arulkumaran K, Sengupta B, Bharath AA. Generative Adversarial Networks: An Overview. IEEE Signal Process Mag 2018;35:53–65. https://doi.org/10.1109/MSP.2017.2765202.

[30] Iqbal T, Ali H. Generative Adversarial Network for Medical Images (MI-GAN). J Med Syst 2018;42:231. https://doi.org/10.1007/s10916-018-1072-9.

[31] Mirza M, Osindero S. Conditional Generative Adversarial Nets 2014.

[32] Cirillo MD, Abramian D, Eklund A. Vox2Vox: 3D-GAN for Brain Tumour Segmentation 2020.

[33] He K, Zhang X, Ren S, Sun J. Identity mappings in deep residual networks. Lect Notes

Comput Sci (Including Subser Lect Notes Artif Intell Lect Notes Bioinformatics) 2016;9908 LNCS:630–45. https://doi.org/10.1007/978-3-319-46493-0_38.

[34] He K, Zhang X, Ren S, Sun J. Deep residual learning for image recognition. Proc. IEEE Comput. Soc. Conf. Comput. Vis. Pattern Recognit., vol. 2016- Decem, 2016, p. 770–8. https://doi.org/10.1109/CVPR.2016.90.

[35] Isola P, Zhu J-Y, Zhou T, Efros AA. Image-to-Image Translation with Conditional Adversarial Networks. 2017 IEEE Conf. Comput. Vis. Pattern Recognit., IEEE; 2017, p. 5967–76. https://doi.org/10.1109/CVPR.2017.632.

[36] Yeung M, Sala E, Schönlieb C-B, Rundo L. Unified Focal loss: Generalising Dice and cross entropy-based losses to handle class imbalanced medical image segmentation 2021.

[37] He K, Zhang X, Ren S, Sun J. Delving Deep into Rectifiers: Surpassing Human-Level Performance on ImageNet Classification 2015.

[38] Demšar J. Statistical comparisons of classifiers over multiple data sets. J Mach Learn Res 2006;7:1–30.

[39] Kingma DP, Ba J. Adam: A Method for Stochastic Optimization 2014.

[40] Loshchilov I, Hutter F. SGDR: Stochastic Gradient Descent with Warm Restarts 2016.

[41] Smith LN, Topin N. Super-Convergence: Very Fast Training of Neural Networks Using Large Learning Rates 2017.

[42] Rela, Munipraveena; Rao SN. Comparative analysis of image enhancement techniques applied to CT liver image. Int J Eng Technol 2018;7:285–9.

[43] Selvaraju RR, Cogswell M, Das A, Vedantam R, Parikh D, Batra D. Grad-CAM: Visual Explanations from Deep Networks via Gradient-Based Localization. Int J Comput Vis 2020;128:336–59. https://doi.org/10.1007/s11263-019-01228-7.

[44] Jonas DE, Reuland DS, Reddy SM, Nagle M, Clark SD, Weber RP, et al. Screening for Lung Cancer With Low-Dose Computed Tomography. JAMA 2021;325:971. https://doi.org/10.1001/jama.2021.0377.

[45] Chung J, Coutinho T, Chu MWA, Ouzounian M. Sex differences in thoracic aortic disease: A review of the literature and a call to action. J Thorac Cardiovasc Surg 2020;160:656–60. https://doi.org/10.1016/j.jtcvs.2019.09.194.

[46] Martin C, Sun W, Primiano C, McKay R, Elefteriades J. Age-Dependent Ascending Aorta Mechanics Assessed Through Multiphase CT. Ann Biomed Eng 2013;41:2565–74. https://doi.org/10.1007/s10439-013-0856-9.

[47] Stoll S, Sowah SA, Fink MA, Nonnenmacher T, Graf ME, Johnson T, et al. Changes in

aortic diameter induced by weight loss: The HELENA trial- whole-body MR imaging in a dietary intervention trial. Front Physiol 2022;13. https://doi.org/10.3389/fphys.2022.976949.

[48] Cayne NS, Veith FJ, Lipsitz EC, Ohki T, Mehta M, Gargiulo N, et al. Variability of maximal aortic aneurysm diameter measurements on CT scan: significance and methods to minimize. J Vasc Surg 2004;39:811–5. https://doi.org/10.1016/j.jvs.2003.11.042.

[49] Chen C-K, Chou H-P, Guo C-Y, Chang H-T, Chang Y-Y, Chen I-M, et al. Interobserver and intraobserver variability in measuring the tortuosity of the thoracic aorta on computed tomography. J Vasc Surg 2018;68:1183-1192.e1. https://doi.org/10.1016/j.jvs.2018.01.047.

[50] Wever JJ, Blankensteijn JD, van Rijn JC, Broeders IAMJ, Eikelboom BC, Mali WPTM. Inter- and Intraobserver Variability of CT Measurements Obtained After Endovascular Repair of Abdominal Aortic Aneurysms. Am J Roentgenol 2000;175:1279–82. https://doi.org/10.2214/ajr.175.5.1751279.

[51] Duquette AA, Jodoin P-M, Bouchot O, Lalande A. 3D segmentation of abdominal aorta from CT-scan and MR images. Comput Med Imaging Graph 2012;36:294–303. https://doi.org/10.1016/j.compmedimag.2011.12.001.

[52] Yu Y, Gao Y, Wei J, Liao F, Xiao Q, Zhang J, et al. A Three-Dimensional Deep Convolutional Neural Network for Automatic Segmentation and Diameter Measurement of Type B Aortic Dissection. Korean J Radiol 2021;22:168. https://doi.org/10.3348/kjr.2020.0313.

[53] Satriano A, Guenther Z, White JA, Merchant N, Di Martino ES, Al-Qoofi F, et al. Three-dimensional thoracic aorta principal strain analysis from routine ECG-gated computerized tomography: feasibility in patients undergoing transcatheter aortic valve replacement. BMC Cardiovasc Disord 2018;18:76. https://doi.org/10.1186/s12872-018-0818-0.

[54] Tian Q, Li X, Li J, Cheng Y, Niu X, Zhu S, et al. Image quality improvement in low-dose chest CT with deep learning image reconstruction. J Appl Clin Med Phys 2022;23. https://doi.org/10.1002/acm2.13796.

[55] Fantazzini A, Esposito M, Finotello A, Auricchio F, Pane B, Basso C, et al. 3D Automatic Segmentation of Aortic Computed Tomography Angiography Combining Multi-View 2D Convolutional Neural Networks. Cardiovasc Eng Technol 2020;11:576–86. https://doi.org/10.1007/s13239-020-00481-z.

[56] Cho J, Lee K, Shin E, Choy G, Do S. How much data is needed to train a medical image deep learning system to achieve necessary high accuracy? 2015.

[57] Comelli A, Dahiya N, Stefano A, Benfante V, Gentile G, Agnese V, et al. Deep learning approach for the segmentation of aneurysmal ascending aorta. Biomed Eng Lett 2021;11:15–24. https://doi.org/10.1007/s13534-020-00179-0.

[58] Paszke A, Chaurasia A, Kim S, Culurciello E. ENet: A Deep Neural Network Architecture for Real-Time Semantic Segmentation 2016.

[59] Xie Y, Padgett J, Biancardi AM, Reeves AP. Automated aorta segmentation in low-dose chest CT images. Int J Comput Assist Radiol Surg 2014;9:211–9. https://doi.org/10.1007/s11548-013-0924-5.

[60] Isola P, Zhu J-Y, Zhou T, Efros AA. Image-to-Image Translation with Conditional Adversarial Networks 2016.

[61] Diakogiannis FI, Waldner F, Caccetta P, Wu C. ResUNet-a: a deep learning framework for semantic segmentation of remotely sensed data 2019. https://doi.org/10.1016/j.isprsjprs.2020.01.013.

[62] Oktay O, Schlemper J, Folgoc L Le, Lee M, Heinrich M, Misawa K, et al. Attention U-Net: Learning Where to Look for the Pancreas 2018.

[63] Chollet F. Xception: Deep Learning with Depthwise Separable Convolutions 2016.

[64] de Menezes LC, de Araujo ARVF, Conci A. An approach based on image processing techniques to segment lung region in chest X-ray images. 2021 34th SIBGRAPI Conf. Graph. Patterns Images, IEEE; 2021, p. 113–20. https://doi.org/10.1109/SIBGRAPI54419.2021.00024.

[65] Raffort J, Adam C, Carrier M, Ballaith A, Coscas R, Jean-Baptiste E, et al. Artificial intelligence in abdominal aortic aneurysm. J Vasc Surg 2020;72:321-333.e1. https://doi.org/10.1016/j.jvs.2019.12.026.

[66] Shum J, Martufi G, Di Martino E, Washington CB, Grisafi J, Muluk SC, et al. Quantitative Assessment of Abdominal Aortic Aneurysm Geometry. Ann Biomed Eng 2011;39:277–86. https://doi.org/10.1007/s10439-010-0175-3.

[67] Quintana RA, Taylor WR. Introduction to the Compendium on Aortic Aneurysms. Circ Res 2019;124:470–1. https://doi.org/10.1161/CIRCRESAHA.119.314765.